\title{The survival probability of diffusion with killing}
\author {David Holcman\thanks{Department of Mathematics, Weizmann
Institute of Science, Rehovot 76100, Israel. D. H.  incumbent to the
Madeleine
 Haas Russell Career Development Chair. },\thanks{Departement de Mathematique et de Biologie, 46, rue d'Ulm
75005 Paris, France.} \ Avi Marchewka\thanks{Physics Department,
University of North Carolina, Raleigh, NC, USA. },\ and Zeev
Schuss\thanks{{Department of Mathematics, Tel-Aviv University,
Ramat-Aviv 69978,  Israel.} }}
\documentclass[12pt,date]{article}
\usepackage{float}
   \usepackage{amsmath}
\usepackage{epsfig, graphicx} %Work with figures.
\newcommand{\mb}[1]{\mbox{\boldmath$#1$}}
\newcommand{\p}{\partial}
\newcommand{\ds}{\displaystyle}
\newcommand{\beq}{\begin{eqnarray}}
\newcommand{\beqq}{\begin{eqnarray*}}
\newcommand{\eeq}{\end{eqnarray}}
\newcommand{\eeqq}{\end{eqnarray*}}
\newcommand{\x}{\mbox{\boldmath$x$}}
\newcommand{\y}{\mbox{\boldmath$y$}}

\font\bb=msbm10 at 12pt

\def\rR{\hbox{\bb R}} 
 
\def\zZ{\hbox{\bb Z}}

\def\ds#1{\displaystyle{#1}}

\def\QED{\quad\hbox{\hskip 4pt\vrule width 5pt height 6pt depth 1.5pt}}

\begin{document}
%%%%%%%%%%%%%%%%%%%%%%%%%%%%%%%%%%%%%%%%%%%%%%%%%%%%%%%%%%%%%%%%
\numberwithin{equation}{section}

\maketitle
\begin{abstract}
We present a general framework to study the effect of killing
sources on moving particles, trafficking inside biological cells. We
are merely concerned with the case of spine-dendrite communication,
where the number of calcium ions, modeled as random particles is
regulated across the spine microstructure by pumps, which play the
killing role. In particular, we study here the survival probability
of ions in such environment and we present a general theory to
compute  the ratio of the number of absorbed particles at specific
location to the number of killed particles during their sojourn
inside a domain. In the case of a dendritic spine, the ratio is
computed in terms of the survival probability of a stochastic
trajectory in a one dimensional approximation. We show that the
ratio depends on the distribution of killing sources.
The biological conclusion follows:  changing the position of the
pumps
is enough to regulate the calcium ions and thus the spine-dendrite
communication.
\end{abstract}

%\newpage

%\tableofcontents
\newpage
%%%%%%%%%%%%%%%%%%%%%%%%%%%%%%%%%%%%%%%%%%%%%%%%%%%%%%%%%%%%%%%%
\section{Introduction}
\begin{figure}%\label{spineHead}
\centering \psfig{figure =  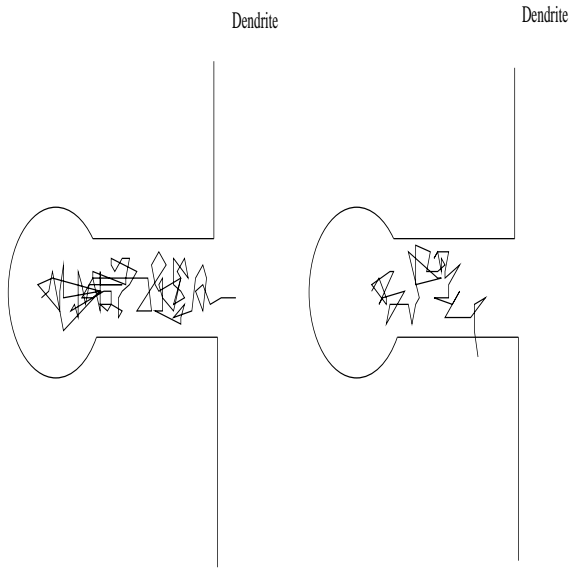,height =75mm,width=75mm}
\caption{\small {\bf Extrusion of an ion from the spine neck}. A
dendritic spine is a microstructure located on the dendrite of
neurons, consisting of a round head connected to the dendrite
through a cylindrical neck. Its function is still unclear. After
ions enter through the head, either they are pumped out (right
figure) or they reach the dendrite (left picture). The number of
ions reaching the dendrite is regulated by the number and the
distribution of pumps and the length of the spine neck. The neck
length changes dynamically and this is induced by previous calcium
ions.} \label{spine1}
\end{figure}

%%%%%%%%%%%%%%%%%%%%%%%%%%%%%%%%%%%%%%%%%%%%%%%%%%%%%%%%%%%%%%%%
The post-synaptic part of a synapse is usually a dendritic spine, a
microstructure located on a dendrite of a neuron (see figure
\ref{spine1}) \cite{Sabatini}, \cite{KS}, \cite{Nicoll}. The spine
geometry consists of a nearly spherical head connected to the
dendrite by a narrow cylindrical neck. Calcium ions enter the spine
head through glutamate gated channels following the release of
glutamate neurotransmitters by the pre-synaptic terminal. The
communication between a dendritic spine and the dendrite depends on
the ability of the calcium ions to pass through the cylindrical neck
to the dendrite. When ions enter the neck, they diffuse and either
reach the dendrite or are extruded on their way to the dendrite by
pump proteins located on the lateral surface of the neck
\cite{HSK,Sabatini}.

The number of calcium ions that arrive at the dendrite and the
calcium contents of the spine are regulated by the geometry of the
neck and by the contents of the spine. The contents include
organelles, such as the endoplasmic reticulum, calcium buffer
proteins such as calmodulin, calcium stores, actin-myosin proteins,
and pumps on the spine membrane. In this paper, we focus on the role
of the spine neck in spine-dendrite communication, which is an area
of intense experimental research (see, e.g., \cite{Sabatini},
\cite{KS}).

We adopt a simplified one-dimensional model of the diffusive motion
of calcium ions in the neck, in which the termination of ionic
trajectories by pumps is described as "killing", while termination
in the dendrite is described as "absorption". A killing measure is
the probability per unit time and unit length to terminate a
trajectory at a given point at a given time. Thus an ion can pass
through a killing site many times without being terminated. In
contrast, an absorbing boundary terminates the trajectory with
probability 1 the first time the trajectory gets there. Thus we
distinguish between two random times on a trajectory, the time to be
killed, denoted $T$, and the time to be absorbed, denoted $\tau$.

We need to find the probability $\Pr\left\{\tau>T\,|\,\y\right\}$ of
an ion getting killed (pumped out) in the neck before it is absorbed
at the boundary (the dendrite), given that it started at a point
$\y$ in the neck. The ratio
 \beqq
 R_{\infty}=\frac{\Pr\left\{\tau<T\,|\,\y\right\}}{\Pr\left\{\tau>T\,|\,\y\right\}}
 \eeqq
is the fraction of absorbed to killed (pumped) particles. We also
need to calculate $E\left[T\,|\,\tau>T,\y\right]$, the mean time to
be killed, given that the particle is killed, as well as
$E\left[\tau\,|\,T>\tau,\y\right]$, the mean time to absorption,
given that the particle is absorbed.

An application of our model in neurobiology concerns calcium
regulation in the dendritic spine and in the dendrite. In dendrites
of neurons, ions are constantly exchanged between compartments and
when the concentration of calcium ions in the dendritic shaft rises
above a threshold value, some specific cascades of chemical
reactions are initiated, that can lead to a new physiological stage,
where the synaptic properties are modified. For example, the
biophysical properties of the synapses or the number of channel
receptors can be irreversibly changed
\cite{Nicoll,choquet1,choquet2}. The process that consists of
changing the synaptic properties is known as synaptic plasticity.
Today, the mechanisms of induction of synaptic changes are still
unclear, but it has been demonstrated recently \cite{KHS} that the
induction process can be affected by the dynamics of the
spine-dendrite coupling. The communication between a dendritic spine
and the dendrite depends on the ability of the ions to pass through
the cylindrical neck of the spine (see figure \ref{spine1}). The
measure of this ability is the parameter $R_\infty$. When ions leave
the spine head and enter the neck, they diffuse and either reach the
dendrite (with probability $\Pr\left\{T>\tau\,|\,\y\right\}$), or,
as mentioned above, are extruded by pump proteins on their way to
the dendrite \cite{HSK,Sabatini}.

In a simplified homogenized model proposed in \cite{KHS},  the
number of ions filtered by the neck has been estimated  and compared
with experimental results. This number depends on the distribution
of pumps along the neck and on the efficiency of the pumping
process. The precise comparison with experimental data in \cite{KHS}
made it possible to predict that changing the length of the spine
neck (which occurs under specific conditions, see for example
\cite{KS}) is sufficient to regulate precisely the number of ions
arriving at the dendrite. Spine-dendrite calcium signaling
(\cite{HSK,Sabatini}) and its regulation through specific
microstructures, such as the spine neck, is crucial for the
induction of synaptic plasticity, which underlies learning and
memory.

The mean time $E[t\,|\,\y]$ an ion spends inside the neck can be
written as
 \beqq
 E[t\,|\,\y]&=&E[t\,|\,\tau<T,\y]\Pr\{\tau<T\,|\,\y\}+E[t\,|\,T<\tau,\y]\Pr\{T<\tau\,|\,\y\}\\
 &&\\
 &=&
 E[\tau\,|\,\tau<T,\y]\Pr\{\tau<T\,|\,\y\}+E[T\,|\,T<\tau,\y]\Pr\{T<\tau\,|\,\y\}.
 \eeqq
The rate $1/E[t\,|\,\y]$ is the total probability flux out of the
neck. This is a measurable quantity that can be used to prove that
ions diffusing into the dendrite originate in the spine head.
Indeed, calcium that enters the spine head through the glutamate
gated channels at the top of the spine head takes much longer to
reach the dendrite than calcium that enters through voltage gated
channels. This is due to the much faster propagation of the membrane
depolarization than movement by diffusion.

In a biological context the final distribution of particles between
absorption and killing indicates the future changes in the steady
properties of the synapse. This is a general principle in cell
biology regulation. It is fundamental for the homeostasis of a
living cell to regulate the number of proteins or small molecules it
contains and to maintain this number constant in the absence of
external input. This is for example achieved through an equilibrium
between synthesis and hydrolysis mechanisms. At a molecular level,
when molecules reach the active sites of free enzymes by a Brownian
random walk,  either the molecules are hydrolyzed or nothing happens
(see \cite{StochasticMM} for a stochastic description) and after
some time, the molecules are absorbed or enter different organelles.
This is what happens in signal transduction, as in synapses of
neurons or in sensor cells. In some cases, the stability and the
function of the cell depends on the efficiency of such dynamical
processes. In addition, the geometry of the cell participates in the
regulation of the number of particles, such as ions, that reach
specific locations.
%%%%%%%%%%%%%%%%%%%%%%%%%%%%%%%%%%%%%%%%%%%%%%%%%%%%%%%%%%%%%%%%%%%%%
%%%%%%%%%%%%%%%%%%%%%%%%%%%%%%%%%%%%%%%%%%%%%%%%%%%%%%%%%%%%%%%%%%%%%%%%%%%%

In the present  work, we are interested in estimating the
probability that an ion survives in a medium containing many pumps.
We compute the probability to arrive at a specific location before
being killed (see figure \ref{spine1}) as a function of structure
and pump distribution. In the case of a dendritic spine, we assume
that the cylindrical neck can be approximated by a one dimensional
interval, and the computations are given in a one-dimensional model.
The one-dimensional approximation is valid when the radius of the
spine neck is sufficiently small, otherwise, the small pumps cannot
affect the normal diffusion process (see \cite{1D}). We will see
that various pump distributions affect the concentration of ions in
the neck; we compare a uniform distribution along the spine neck,
modeled as a constant killing rate, to an accumulation of pumps in
``hot spots'' at some specific locations, (for example at the base
of the dendritic spine). In either case, we estimate the flux of
ions into the dendrite.

The reduced one-dimensional model of Brownian motion with killing
and absorption is investigated in various types of killing sets. It
is of interest to determine the influence of spatial  distribution
of the killing measure on the global survival probability of the
population. Absorption and killing are expressed differently in the
Fokker-Planck equation (FPE) for the transition probability density
function (pdf) of the Brownian motion. While total absorption at the
boundary is expressed as a homogeneous Dirichlet boundary condition,
killing appears as a reaction term in the FPE \cite{book}.

Our main results are general expressions for the probabilities,
ratio, and mean times in general, and in particular, we give
explicit expressions as functions of the geometry and distribution
of killing sites in the one-dimensional model. We also provide a
biological interpretation of the results.

%%%%%%%%%%%%%%%%%%%%%%%%%%%%%%%%%%%%%%%%%%%%%%%%%%%%%%%%%%%%%%%%
\section{Killing measure and the survival probability}
%%%%%%%%%%%%%%%%%%%%%%%%%%%%%%%%%%%%%%%%%%%%%%%%%%%%%%%%%%%%%%%%
We consider a Brownian motion (particle) in a cylinder, whose
lateral boundary contains many small absorbing hole, one base is
reflecting and the other absorbing. This model can be approximated
\cite{1D} by a one-dimensional Brownian motion on an interval with
one reflecting and one absorbing endpoints, and a killing measure
inside the interval. The strength of the killing measure is related
to the absorption flux of the three-dimensional Brownian motion
through the small holes on the boundary of the cylinder. The killing
measure $k(x,t)$ is the probability per unit time and unit length
that the Brownian trajectory is terminated at point $x$ and time $t$
\cite{book}.

The survival probability and the pdf of the surviving trajectories
can be derived from the Wiener measure \cite{ItoMcKean}. Indeed, for
a Brownian trajectory $X(t)$ and the random time at which it is
terminated, $\tau $, we denote the (defective) probability density
of finding a trajectory at point $x$, given that it starts at $y$,
by
\begin{eqnarray*}
u(x,t\,|\,y)\,dx=\Pr \left\{x(t) \in x+dx, \,\tau>t\,|\,x(0)
=y\right\} .
\end{eqnarray*}
To derive the joint density of $x(\tau)$ and $\tau $, we can
formulate the problem in terms of the Wiener integral with a
killing measure. The Wiener density per unit time of being killed
in the time interval $[t,t+\Delta t]$ at a point $x_{N}=x$ is
\begin{eqnarray}
&&\Pr \left\{x(\tau) =x,\,\tau=t\,|\,x(0) =y\right\} =\label{Wiener} \\
&&\nonumber\\
&&\lim_{N\rightarrow \infty }\frac{1}{\Delta t}k\left(
x_{N},t_{N}\right) \Delta t{\Bigg[} \left( \frac{1}{2\pi \Delta
t}\right) ^{N/2}
\underbrace{\int\cdots\int}_{N}u_{0}(x_0)\times\nonumber\\
&&\nonumber\\
&&\prod_{j=1}^N {\Bigg(}\exp\left\{-\frac{\left(
x_j-x_{j-1}\right)^2}{2\Delta t}\right\}
\left[1-k\left(x_{j-1},t_{j-1}\right)\Delta
t\right]\,dx_{j-1}{\Bigg) }
{\Bigg]}\nonumber\\
&&\nonumber\\
&&=k\left( x,t\right) u\left( x,t\,|\,y\right) ,\nonumber
\end{eqnarray}
where
\begin{eqnarray*}
\Delta t=\frac{t}{N},\quad t_{j}=j\Delta t,
\end{eqnarray*}
and $u\left( x,t\,|\,y\right) $ is the solution of the initial
value problem
\begin{eqnarray}
u_{t} &=&u_{xx}-k\left( x,t\right) u,\quad\mbox{for $x\in\rR,\ t>0$}  \nonumber \\
&&\label{FPE}\\
 u\left( x,0\right)  &=&\delta \left( x-y\right) .
\nonumber
\end{eqnarray}
In the case that $k\left( x,t\right) =V_{0}$ and  the diffusion
coefficient is $D$, we have
\begin{eqnarray}
 \frac{\partial u(x,t\,|\,y)}{\partial t}&=& D\frac{\partial^{2} u(x,t\,|\,y)}
{\partial x^{2}} -V_{0}u(x,t\,|\,y),\quad \mbox{for $x\in\rR,\ t>0$} \nonumber\\
&&\label{first}\\
u(x,0\,|\,y)& =& \delta(y-x). \nonumber
\end{eqnarray}
The solution is given by
\begin{eqnarray}
u(x,t\,|\,y)=\frac{1}{2\sqrt{\pi Dt}}
\exp\left\{-V_{0}t-\frac{(x-y)^{2}}{4Dt}\right\}.\label{k=V0}
\end{eqnarray}
The effect of absorption is expressed through many different
features of the Wiener integral. First, the probability per unit
time of being killed (absorbed) inside the interval $\left[
a,b\right] $ at time $t$ is
\begin{eqnarray*}
\Pr \left\{ x(\tau)\in[a,b],\,\tau =t\,|\,x(0) =y\right\}
&=&\int_{a}^{b}k\left( x,t\right) u\left( x,t\,|\,y\right) \,dx,
\end{eqnarray*}
while the probability of being killed in the interval before time
$t$ is
\begin{eqnarray*}
\Pr \left\{x(\tau) \in[a,b],\,\tau <t\,|\,x(0) =y\right\}
&=&\int_{0}^{t}\int_{a}^{b}k(x,t)u(x,t\,|\,y)\,dx\,dt.
\end{eqnarray*}
The probability of ever being killed in the interval is
\begin{eqnarray*}
\Pr \left\{ x(\tau) \in \left[ a,b\right] \,|\,x(0) =y\right\}
&=&\int_{0}^{\infty }\int_{a}^{b}k\left( x,t\right) u\left(
x,t\,|\,y\right) \,dx\,dt,
\end{eqnarray*}
and the density of ever being killed at $x$ is therefore
\begin{eqnarray}
\Pr \left\{ x(\tau) =x\,|\,\,x(0) =y\right\} &=&\int_{0}^{\infty
}k(x,t) u\left( x,t\,|\,y\right) \,dt .\label{everatx}
\end{eqnarray}
The survival probability is the probability that the trajectory
still exists at time $t$, that is,
\begin{eqnarray*}
S(t)=\Pr \{\tau >t|\,x(0)=y\}=\int_{\rR} u(x,t\,|\,y)\,dx.
\end{eqnarray*}
 For the case  $k(x,t) =V_{0}$ eq.(\ref{k=V0}) gives
\begin{eqnarray}
\Pr \{\tau >t\,|\,x(0) =y \}=\int_{\rR} u(x,t\,|\,y)\, dx =
e^{-V_{0}t}.
\end{eqnarray}
This is exactly the rate at which particles disappear from the
medium. The rate is exponential, so that out of $N_0$ initial
independent Brownian particles in $\rR$ the expected number of
particles that have disappeared by time $t$  is
$N_{0}(1-e^{-V_{0}t})$. The probability of being killed at point
$x$, given by eq.(\ref{everatx}), is
\begin{eqnarray}
P(x\,|\,y)&=&V_{0}\int_{0}^\infty \frac{1}{2\sqrt{\pi Dt}}
\exp\left\{-V_{0}t-\frac{(x-y)^{2}}{4Dt}\right\}\, dt\nonumber\\
&&\label{everatxV0}\\
&=&\frac12\sqrt{\frac{V_0}{D}}\exp\left\{-\displaystyle{\sqrt{\frac{V_{0}}{D}}}
| x-y|\right\}.\nonumber
\end{eqnarray}
We assume henceforward that the killing measure is time independent.
%%%%%%%%%%%%%%%%%%%%%%%%%%%%%%%%%%%%%%%%%%%%%%%%%%%%%%%%%%%%%%%%
\section{Absorption versus killing}
%%%%%%%%%%%%%%%%%%%%%%%%%%%%%%%%%%%%%%%%%%%%%%%%%%%%%%%%%%%%%%%%
We consider now a particle diffusing in a domain
$\Omega\subset\rR^n$ with a killing measure $k(\x)$ and an absorbing
part $\p\Omega_a\subset\p\Omega$ of the boundary $\p\Omega$. Thus
the trajectory of the particle can terminate in two ways, it can
either be killed inside $\Omega$ or absorbed in $\p\Omega_a$. The
difference between the killing and the absorbing processes is that
while the trajectory has a finite probability of not being
terminated at points $\x$ where $k(\x)>0$, it is terminated with
probability 1 the first time it hits $\p\Omega_a$. Thus the
trajectory may traverse many times killing regions, where $k(\x)>0$,
but it cannot emerge from the absorbing part of the boundary.

%%%%%%%%%%%%%%%%%%%%%%%%%%%%%%%%%%%%%%%%%%%%%%%%%%%%%%%%%%%%%%%%
\subsection{Definition and basic equations}
%%%%%%%%%%%%%%%%%%%%%%%%%%%%%%%%%%%%%%%%%%%%%%%%%%%%%%%%%%%%%%%%
We define two random termination times defined on the trajectories
of the diffusion process: the time to killing, denoted $T$, and the
time to absorption in $\p\Omega_a$, denoted $\tau$, which is the
first passage time to $\p\Omega_a$.  We calculate below the
probability $\Pr\{T<\tau\,|\,y\}$, and the conditional distribution
$\Pr\{\tau<t\,|\,\tau<T,\,y\}$.

We consider the trajectories of the stochastic differential
equation
 \beq
d{\x}=\mb{a}(\x)\,dt+\mb{B}(\x)\,d{\mb{w}}(t)\quad\mbox{for
$\x(t)\in\Omega$},\label{sde}
 \eeq
where $\mb{a}(\x)$ is a smooth drift vector, $\mb{B}(\x)$ is a
smooth diffusion matrix, and $\mb{w}(t)$ is a vector of
independent standard Brownian motions \cite{book}. We assume that
a killing measure $k(\x)\geq0$ is defined in $\Omega$ and
$k(\x)>0$ on a set of positive measure.

The transition probability function of $\x(t)$ satisfies the
Fokker-Planck equation
 \beq
 \frac{\p p(\x,t\,|\,\y)}{\p t}={\cal
 L}p(\x,t\,|\,\y)-k(\x)p(\x,t\,|\,\y)\quad\mbox{for}\quad \x,\y\in\Omega,\label{FPEp}
 \eeq
where the forward operator ${\cal L}$ is defined by
 \beq
{\cal
L}p(\x,t\,|\,\y)=\sum_{i,j=1}^n\frac{\p^2\sigma^{i,j}(\x)p(\x,t\,|\,\y)}{\p
x^i\p x^j} -\sum_{i=1}^n\frac{\p a^i(\x)p(\x,t\,|\,\y)}{\p
x^i},\label{FO}
 \eeq
and
 \[\mb{\sigma}(\x)=\frac{1}2\mb{B}(\x)\mb{B}^T(\x).\]
The forward operator ${\cal L}$ can also be written in the
divergence form
 \beq
{\cal L}p(\x,t\,|\,\y)=-\nabla\cdot\mb{J}(\x,t\,|\,\y),
 \eeq
where the components of the flux density vector
$\mb{J}(\x,t\,|\,\y)$ are defined as
 \beq
J^i(\x,t\,|\,\y)=-\sum_{j=1}^n\frac{\p\sigma^{i,j}(\x)p(\x,t\,|\,\y)}{\p
x^i}+a^i(\x)p(\x,t\,|\,\y).\label{Ji}
 \eeq
The initial and boundary conditions for the Fokker-Planck equation
(\ref{FPEp}) are
 \beq
 p(\x,0\,|\,\y)&=&\delta(\x-\y)\quad\mbox{for}\quad\x,\y\in\Omega\label{IC}\\
 &&\nonumber\\
 p(\x,t\,|\,\y)&=&0\quad\mbox{for}\quad t>0,\ \x\in\p\Omega,\ \y\in\Omega_a\label{BC}\\
 &&\nonumber\\
\mb{J}(\x,t\,|\,\y)\cdot\mb{\nu}(\x)&=&0\quad\mbox{for}\quad t>0,\
\x\in\p\Omega-\p\Omega_a,\ \y\in\Omega.\label{noflux}
 \eeq
The transition pdf $p(\x,t\,|\,\y)$ is actually the joint pdf
 \beq
 p(\x,t\,|\,\y)\,d\x=\Pr\{\x(t)\in\x+d\x,\,T>t,\,\tau>t\,|\,\y\},\label{pxTtau}
 \eeq
that is, $p(\x,t\,|\,\y)$ is the probability density that the
trajectory survived to time $t$, i.e., was neither killed nor
absorbed in $\p\Omega_a$, and is located at $\x$.

We begin by showing that
 \beq
 \Pr\{T<\tau\,|\,\y\}=\int_0^\infty\int_{\Omega}
 k(\x)p(\x,t\,|\,\y)\,d\x\,dt\label{Ttau}
 \eeq
by two different derivations. First, assume that the entire
boundary is absorbing, that is, $\p\Omega_a=\p\Omega$. Then the
probability density of surviving up to time $t$ and being killed
at time $t$ at point $\x$  can be represented by the limit as
$N\to\infty$ of
 \beq
&&\Pr\Big{\{}{\x}_N(t_{1,N})\in\Omega,{\x}_N(t_{2,N})\in\Omega,\dots,
{\x}_N(t)=\x, t\leq T\leq t+\Delta t\,|\,\x(0)=\y\Big{\}}\nonumber=\\
&&\nonumber\\
&&\Bigg{[}\int_{\Omega} \int_{\Omega}\cdots
\int_{\Omega}\,\prod_{j=1}^{N} \frac{d{\y}_j}{\sqrt{(2\pi \Delta
t)^n\det\mb{\sigma}(\x)(t_{j-1,N}))}}
\nonumber\\
&&\nonumber\\
 &&\times \exp \Bigg{\{} -\frac{1}{2\Delta t}
\left[\mb{\y}_j-\x(t_{j-1,N})- \mb{a}({\x}(t_{j-1,N}))\Delta t
\right]^T\mb{\sigma}^{-1}(\x(t_{j-1,N}))
\nonumber\\
&&\nonumber\\
&&\times\left[{\y}_j-\x(t_{j-1,N})-\mb{a}(\x(t_{j-1,N}))\Delta t
\right]\Bigg{\}}\left[1-k(\x(t_{j,N})\Delta
t\right]\Bigg{]}k(\x)\Delta t ,\label{Wiener1}
 \eeq
 where
\begin{eqnarray*}
\Delta t=\frac{t}{N},\quad t_{j,N}=j\Delta t,
\end{eqnarray*}
and
\[\x(t_{0,N})=\y\]
in the product. The limit is the Wiener integral defined by the
stochastic differential equation (\ref{sde}), with the killing
measure $k(\x)$ and the absorbing boundary condition
\cite{Kleinert}. In the limit $N\to\infty$ the integral
(\ref{Wiener1}) converges to the solution of the Fokker-Planck
equation (\ref{FPEp}) in $\Omega$ with the initial and boundary
conditions (\ref{IC}) and (\ref{BC}). Integrating over $\Omega$ with
respect to $\x$ and from $0$ to $\infty$ with respect to $t$, we
obtain, in view of (\ref{pxTtau}), the representation (\ref{Ttau}).

A second derivation begins with the integration of the
Fokker-Planck equation (\ref{FPEp}),
 \beq
1=\int_0^\infty\oint_{\partial\Omega}\mb{J}(\x,t\,|\,\y)\cdot\mb{\nu}
(\x)\,dS_{\x}\,dt+\int_0^\infty\int_{\Omega}k(\x)p(\x,t\,|\,\y)\,d\x\,dt.\label{IFPE}
 \eeq
We write
 \beq
 J(t\,|\,\y)=\oint_{\partial\Omega}\mb{J}(\x,t\,|\,\y)\cdot\mb{\nu}(\x)\,dS_{\x}
 \eeq
and note that this is the absorption probability current on
$\p\Omega$. Therefore, in view of the boundary conditions
(\ref{BC}), (\ref{noflux}), $\ds\int_0^\infty J(t\,|\,\y) \,dt$ is
the total probability that has ever been absorbed at the boundary
$\p\Omega_a$. This is the probability of trajectories that have
not been killed before reaching $\p\Omega_a$. Writing
eq.(\ref{IFPE}) as
 \beqq
 \int_0^\infty J(t\,|\,\y)
 \,dt=1-\int_0^\infty\int_{\Omega}k(\x)p(\x,t\,|\,\y)\,d\x\,dt,
 \eeqq
we obtain (\ref{Ttau}).

The probability distribution function of $T$ for trajectories that
have not been absorbed in $\p\Omega_a$ is found by integrating the
Fokker-Planck equation with respect to $\x$ over $\Omega$ and with
respect to $t$ from $0$ to $t$. It is given by
 \beq
\Pr\{T<t\,|\,\tau>T,\y\}&=&\frac{\Pr\{T<t,\tau>T\,|\,\y\}}
{\Pr\{\tau>T\,|\,\y\}}\nonumber\\
&&\nonumber\\
&=&\frac{\ds\int_0^t\int_{\Omega}k(\x)p(\x,s\,|\,\y)\,d\x\,ds}{\ds\int_0^\infty
\int_{\Omega}k(\x)p(\x,s\,|\,\y)\,d\x\,ds}.
 \eeq
Hence
 \beq
 E[T\,|\,T<\tau,\y]=\frac{\ds\int_0^\infty\int_t^\infty\int_{\Omega}k(\x)p(\x,s\,|\,\y)
 \,d\x\,ds\,dt}{\ds\int_0^\infty
\int_{\Omega}k(\x)p(\x,s\,|\,\y)\,d\x\,ds}.
 \eeq
Equivalently,
 \beq
 E[T\,|\,T<\tau,\y]=\frac{\ds\int_0^\infty s\int_{\Omega}k(\x)p(\x,s\,|\,\y)
 \,d\x\,ds\,dt}{\ds\int_0^\infty
\int_{\Omega}k(\x)p(\x,s\,|\,\y)\,d\x\,ds},
 \eeq
which can be expressed in terms of the Laplace transform
 $$\hat{p}(\x,q\,|\,\y) = \ds\int_0^\infty p(\x,s\,|\,\y)e^{-qs}ds$$ as
\beq
 E[T\,|\,T<\tau,\y]&=&-\frac{\ds \int_{\Omega}k(\x)\hat{p}'(\x,q\,|\,\y)
 \,d\x}{\ds \int_{\Omega}k(\x)\hat{p}(\x,q\,|\,\y)d\x}\nonumber\\
 &&\label{cmfptlp}\\
 &=&-\left. \frac{\p}{\p q} \left(\ln \left\{\int_{\Omega}k(\x)\hat{p}(\x,q\,|\,\y)d\x
 \right\}\right)\right|_{q=0}.\nonumber
 \eeq
The conditional distribution of the first passage time to the
boundary of trajectories, given they are absorbed, is
 \beq
 \Pr\{\tau<t\,|\,T>\tau,\y\}=\frac{\ds\int_0^tJ(s\,|\,\y)\,ds}{1-\ds\int_0^\infty
\int_{\Omega}k(\x)p(\x,s\,|\,\y)\,d\x\,ds}.
 \eeq
Thus the mean time to absorption in $\p\Omega_a$ of trajectories
that are absorbed is given by \cite{Direct}
 \beq
 E[\tau\,|\,T>\tau,\,y]&=&\int_0^\infty\Pr\{\tau>t\,|\,T>\tau,y\}\,dt\nonumber\\
 &&\nonumber\\
 &=&\frac{\ds\int_0^\infty s J(s\,|\,y)\,ds}{1-\ds\int_0^\infty
\int_{\Omega}k(\x)p(\x,s\,|\,y)\,d\x\,ds}.
 \eeq

The survival probability is given by
 \beq \label{survii}
 S(t\,|\,y)=\int_\Omega p(\x,t\,|\,y)\,d\x.
 \eeq

%%%%%%%%%%%%%%%%%%%%%%%%%%%%%%%%%%%%%%%%%%%%%%%%%%%%%%%%%%%%%%%%
\subsection{An application: a ``hot spot" killing in a finite interval}
%%%%%%%%%%%%%%%%%%%%%%%%%%%%%%%%%%%%%%%%%%%%%%%%%%%%%%%%%%%%%%%%
We provide here first an explicit estimation of the survival
probability when the killing measure is a Dirac killing at a single
point in a finite interval and second, we estimate the conditional
mean first passage time to exit before being killed.

%%%%%%%%%%%%%%%%%%%%%%%%%%%%%%%%%%%%%%%%%%%%%%%%%%%%%%%%%%%%%%%%
\subsubsection{Explicit decay of the survival probability}
%%%%%%%%%%%%%%%%%%%%%%%%%%%%%%%%%%%%%%%%%%%%%%%%%%%%%%%%%%%%%%%%
To compare the survival probability of Brownian motion with and
without a Dirac killing at a point $x_1$ in the interval $[0,\pi]$
with absorbing boundaries, we consider the solution of the
boundary value problem
\begin{eqnarray}
 \frac{\partial u(x,t\,|\,x_1)}{\partial t}&=&
D\frac{\partial^{2} u(x,t\,|\,x_1)} {\partial x^{2}}
-V\delta(x-x_{1})u(x,t\,|\,x_1)
\hbox{ on } \rR \label{edps}\\
&&\nonumber\\
 u(x,0\,|\,x_1) &=& \delta(x-y).\nonumber\\
 &&\nonumber\\
u(0,t\,|\,x_1) &=& u(\pi,t\,|\,x_1)=0,\nonumber
\end{eqnarray}
and we denote by $G$ the Green function of the free particle
problem, where $V=0$, then
$$G(x,t\,|\,y)=\frac2\pi\sum_{n=1}^\infty \sin nx\sin ny e^{-n^2t}.$$
 Therefore the survival probability of Brownian motion in the interval is
 $$S_0(t\,|\,y)=\int_0^\pi G(x,t\,|\,y)\,dx=
\frac4\pi\sum_{n=1}^\infty\frac{\sin(2n-1)y}{2n-1}e^{-(2n-1)^2t}.$$

Using the Laplace transform,  the solution $u(x,t\,|\,y)$ of
(\ref{edps}) with
 $V>0$ is given by
 \beq
 \hat u_V(x,q\,|\,y)=\hat G(x,q\,|\,y)-\frac{V\hat G(x,q\,|\,x_1)}
 {1+V\hat G(x_1,q\,|\,x_1)}\hat G(x_1,q\,|\,y),\label{withdelta}
 \eeq
 where
 \beq
\hat G(x,q\,|\,y)=\frac2\pi\sum_{n=1}^\infty \frac{\sin nx\sin
ny}{q+n^2}.\label{Ghat1}
 \eeq
%The reader is referred to Section \ref{examples} for additional
%details.
Note that
 $$\hat S_0(q\,|\,y)=\int_0^\pi\hat G(x,q\,|\,y)\,dx=
\frac4\pi\sum_{n=1}^\infty
\frac{\sin(2n-1)y}{(2n-1)\left(q+(2n-1)^2\right)}.$$ According to
equation (\ref{survii}), the survival probability $S_V(t\,|\,y)$ is
given by
\begin{eqnarray}
 S_V(t\,|\,y)=\int_0^{\pi} u_V(x,t\,|\,y)\,dx
\end{eqnarray}
and the Laplace transform is
\begin{eqnarray}
 \hat{S}_V(t\,|\,y)=\int_0^{\pi} \hat{u}_V(x,t\,|\,y)\,dx.
\end{eqnarray}
Using (\ref{withdelta}), we find that the survival probabilities,
without and with the Dirac killing, differ by
 \beq \label{rhs}
 \hat S_0(q\,|\,y)- \hat S_V(q\,|\,y)=\frac{V\hat G(x_1,q\,|\,y)}
 {1+V\hat G(x_1,q\,|\,x_1)}\hat S_0(q\,|\,x_1).
 \eeq
To compute $\hat S_0(q\,|\,y)$, we use the formula
\begin{eqnarray*}
Su(q|z)=\sum_1^{\infty} \frac{\cos(nz)}{n^2+q}=\left\{
 \begin{array}{lll}
\ds{\left(\frac{\cosh(\sqrt{q} z)}{\tanh(\sqrt{q} \pi)}
-\frac{1}{\sqrt{q}  \pi}\right)\frac{\pi}{2\sqrt{q}}}, &&
 \mbox{ for } q \geq 0\\
 &&\\
\ds{\left(-\frac{\cos(\sqrt{-q} z)}{\tan(\sqrt{-q} \pi)}
+\frac{1}{\sqrt{-q}  \pi}\right)
\frac{\pi}{2\sqrt{-q}}}, && \mbox{ for } q < 0\\
  &&\\
\end{array}\right.
\end{eqnarray*}
when $x,y \in ]0,\pi[$. Then,
\begin{eqnarray*}
 \hat G(x,q\,|\,y)= \frac{Su(q|x-y)-Su(q|x+y)}{\pi}.
\end{eqnarray*}
Thus
 \beqq
\hat
S_0(q\,|\,y)&=&\int_0^{\pi}\frac{1}{2\sqrt{q}}\left(\frac{\cosh(\sqrt{q}(x-y))}
{\tanh(\sqrt{q} \pi)}-\frac{\cosh(\sqrt{q} (x+y))}{\tanh(\sqrt{q} \pi)}\right)\,dx \\
&&\\
&=&\frac{1}{2{q}\tanh(\sqrt{q}
\pi)}\left[\sinh(\sqrt{q}(x-y))-\sinh(\sqrt{q}(x+y))\right]_0^{\pi}.
 \eeqq
From a Taylor expansion around $q=0$, we obtain that
 \beq
\hat S_0(q\,|\,y)= \frac{Q(y)}{6\pi} +O(\sqrt{q}),
 \eeq
where $ Q(y) = -3\pi^2y-\pi 3y^2+y^3$, and similarly
 \beq \hat
S_0(q\,|\,x_1)= \frac{Q(x_1)}{6\pi} +O(\sqrt{q}).
 \eeq
We conclude that in a bounded interval,  the decay rate for the
survival probability of a free particle is exponential with a rate
constant $\ds{\frac{6\pi}{Q(y)}}$, which depends on the initial
position of the particle and is given by
 \beqq
S_0(t\,|\,y) \sim \exp\left\{- \frac{6\pi
t}{Q(y)}\right\}\quad\mbox{for}\quad t\gg1.
 \eeqq
For a given $V>0$, equation (\ref{rhs}) contains the term
$\ds{\frac{V\hat G(x_1,q\,|\,y)} {1+V\hat G(x_1,q\,|\,x_1)}}$, which
is 1 at $q=0$, and at the first order approximation, when $y \neq
x_1$,
 \beq \hat S_V(q\,|\,y)=\hat
S_0(q\,|\,y)-S_0(q\,|\,x_1)=\frac{Q(y)}{6\pi}-\frac{Q(x_1)}{6\pi}
+O(\sqrt{q}).
 \eeq
We conclude that the strength $V$ does not enter the first
approximation of the survival probability, but the decay rate
constant is bigger than in pure diffusion. More specifically, we
obtain that
 \beq
S_V(t\,|\,y) \sim \exp\left\{- \frac{6\pi t}{Q(y)-Q(x_1)
}\right\}\quad\mbox{for}\quad t\gg1.
 \eeq
The potential strength $V$ enters in the next term in the expansion
of $S_V$.
%%%%%%%%%%%%%%%%%%%%%%%%%%%%%%%%%%%%%%%%%%%%%%%%%%%%%%%%%%%%%%%%
\subsubsection{Computation of the conditional MFPT $ E[T\,|\,T<\tau,y]$ }
%%%%%%%%%%%%%%%%%%%%%%%%%%%%%%%%%%%%%%%%%%%%%%%%%%%%%%%%%%%%%%%%
The conditional MFPT $ E[T\,|\,T<\tau,y]$ is computed by using
expression (\ref{cmfptlp}) as follows. Equation (\ref{cmfptlp}),
corresponding to the killing measure $V\delta(x-x_1)$, gives
 \beq
E[T\,|\,T<\tau,y] = -\left.\frac{\p}{\p q}
\ln\{\hat{p}(x_1,q\,|\,y)\}\right|_{q=0}.
 \eeq
The Laplace transform of equation (\ref{edps}) with absorbing
boundary conditions is given by
 \beq
\hat{u}(x,q\,|\,y) = -\frac{2V}{\pi} \sum_{1}^{+\infty}\frac{\sin
nx\sin ny}{q+n^2}\hat{u}(x_1,q\,|\,y) + \hat G(x,q\,|\,y), \eeq
which gives for $x =x_1$, \beq \hat{u}(x_1,q\,|\,y) = \frac{ \hat
G(x_1,q\,|\,y)}{1+ \ds{\frac{2V}{\pi} \sum_{1}^{\infty}\frac{\sin
nx_1\sin ny}{q+n^2}}},
 \eeq
and
 \beqq
\frac{\p}{\p q} \ln \hat{p}(x_1,q\,|\,y) &=& \frac{\p}{\p q} \ln
\hat G(x_1,q\,|\,y) - \frac{\p}{\p q} \ln\left(1+ \frac{2V}{\pi}
\sum_{1}^{+\infty}\frac{\sin nx_1\sin ny}{q+n^2}\right)\nonumber
\\
&&\\
&=&\alpha(x_1\,|\,y) +\beta(x_1\,|\,y)\nonumber
 \eeqq
  with
\beq
 \alpha(x_1\,|\,y) =\left.\frac{\p}{\p q} \ln  \hat
G(x_1,q\,|\,y)\right|_{q=0}
=-\frac{\ds{\sum_{n=1}^{\infty}\frac{\sin nx_1\sin
ny}{n^4}}}{\ds{\sum_{n=1}^{\infty}\frac{\sin nx_1\sin
ny}{n^2}}}\nonumber
 \eeq
  and
\beqq
 \beta(x_1\,|\,y)&=&-\left.\frac{\p}{\p q} \ln \left(1+ \frac{2V}{\pi}
\sum_{1}^{+\infty}\frac{\sin nx_1\sin
ny}{q+n^2}\right)\right|_{q=0}\\
&&\\
& =& \frac{\ds{\frac{2V}{\pi} \sum_{n=1}^{\infty}\frac{\sin
nx_1\sin ny}{n^4}}}{1+\ds{\frac{2V}{\pi}
\sum_{1}^{+\infty}\frac{\sin nx_1\sin ny}{n^2}}}.\nonumber
 \eeqq
 For $x_1,y \in ]0,\pi[$, it is
well known that
 \beqq
\frac{2}{\pi}\sum_{n=1}^{\infty}\frac{\sin nx_1\sin ny}{n^2}&=&
\frac{ (\pi-x_1)y}{\pi}\\
&&\\
\frac{2}{\pi}\sum_{n=1}^{\infty}\frac{\sin nx_1\sin ny}{n^4}& =&
\frac{x_1y}{6\pi}(x_1^2+y^2+2\pi^2)- \frac{(x_1^3+y^3)}{6},
 \eeqq
so that finally, we obtain
 \beqq
E[T\,|\,T<\tau,y]& =&
-\alpha(x_1\,|y,)+\beta(x_1\,|\,y)\\
&&\\
&=&\frac{{x_1y}(x_1^2+y^2+2\pi^2)- \pi{(x_1^3+y^3)}}{6
(\pi-x_1)y}\, \frac{\pi}{\pi+V(\pi-x_1)y}.
 \eeqq

%%%%%%%%%%%%%%%%%%%%%%%%%%%%%%%%%%%%%%%%%%%%%%%%%%%%%%%%%%%%%%%%
\subsection{Ratio measuring the distribution of particles}
%%%%%%%%%%%%%%%%%%%%%%%%%%%%%%%%%%%%%%%%%%%%%%%%%%%%%%%%%%%%%%%%
According to the Fokker-Planck equation (\ref{FPEp}), the time
dependent ratio $R(t)$ of the absorbed particles (particles leaving
the domain, before being killed) to the killed particles at time $t$
can be defined as
 \beq
R(t) =
\frac{\ds{\int_{\partial\Omega_a}}\mb{J}(\x,t\,|\,\y)\cdot\mb{\nu}(\x)}{\ds{\int_{\Omega}}
k(\x)p(\x,t\,|\,\y)\,d\x}.
 \eeq
More interestingly, we can define a steady state ratio $R_{\infty}$,
which is the total number of absorbed particles to the total number
of killed particles after infinite time, by the expression
 \beq
R_{\infty} = \frac{\displaystyle \int_0^\infty
\displaystyle\int_{\partial\Omega_a}\mb{J}(\x,t\,|\,\y)\cdot\mb{\nu}
(\x)\,dS_{\x}\,dt}{\displaystyle\int_0^\infty
\displaystyle\int_{\Omega}k(\x)p(\x,t\,|\,\y)\,d\x\, dt }= \frac{
\displaystyle\int_{\partial\Omega_a}\mb{J}(\x\,|\,\y)
\cdot\mb{\nu}(\x)\,dS_{\x}}{\displaystyle\int_{\Omega}k(\x)G(\x\,|\,\y)\,d\x
},
 \eeq
where $G(\x \,|\,\y)$ is defined by the equation
 \beq
 -\rho(\y)={\cal  L}G(\x,|\,\y)-k(\x)G(\x \,|\,\y)\quad\mbox{for $\x,\y\in\Omega$}\label{FPEpp}
 \eeq
with the forward operator ${\cal L}$ defined in eq.(\ref{FO}),
$\rho(\y)$ is the initial density, and $\mb{J}(\x\,|\,\y)$ is the
flux density vector at point $\x$, computed with respect to the
function $G(\x \,|\,\y)$. When  $\rho(\y)=\delta(\y)$, $G$ is the
standard Green function with boundary conditions given by equation
(\ref{BC}).

We can define another ratio of interest: in a permanent regime, when
a flux enters the domain through a part of the boundary, it is
partitioned into the flux of absorbed and killed particles. When a
steady state regime is achieved, we can define the ratio $R_s$ as
above. We denote by  $\p \Omega_i$ the part of the boundary, where a
steady flux enters the domain. The steady state Fokker-Planck
equation becomes
 \beq
 0={\cal
 L}p(\x\,|\,\y)-k(\x)p(\x\,|\,\y)\quad\mbox{for
 $\x,\y\in\Omega$}\label{FPEps},
 \eeq
where the forward operator ${\cal L}$ is defined by (\ref{FO}) and
the boundary conditions are
 \beqq
 p(\x\,|\,\y)&=&0\quad\mbox{for  $,\x\in\p\Omega,\y\in\Omega_a$}\label{BCp}\\
 &&\nonumber\\
\mb{J}(\x\,|\,\y)\cdot\mb{\nu}(\x)&=&0 \quad\mbox{for $\x\in\p\Omega-\p\Omega_a-\p \Omega_i,\ \y\in\Omega,\ t>0$}. \\&&\nonumber\\
\mb{J}(\x\,|\,\y)\cdot\mb{\nu}(\x)&=&-\Phi(\x) \quad\mbox{for
$\x\in\p \Omega_i $}. \label{nofluxp}
 \eeqq
The time independent flux is $\Phi(\x)\geq 0$. The external steady
state flux of absorbed particles is
 \beq {J_a} = \int_{\p
\Omega_a} \mb{J}(\x\,|\,\y) \cdot\mb{\nu}(\x) dS_{\x}.
 \eeq
The total inward flux is
 \beq {J_i} = \int_{\p \Omega_i}
\mb{J}(\x\,|\,\y) \cdot\mb{\nu}(\x) dS_{\x}= \int_{\p \Omega_i}
\Phi(\x) dS_{\x}.
 \eeq
We define the ratio $R_s$ as
 \beq
R_s=\frac{\displaystyle\int_{\partial\Omega_a}\mb{J}(\x\,|\,\y)
\cdot\mb{\nu}(\x)\,dS_{\x}}{\displaystyle\int_{\Omega}k(\x)p(\x\,|\,\y)\,d\x
} =\frac{\displaystyle\int_{\p \Omega_i} \Phi(\x)
dS_{\x}-\int_{\Omega}k(\x)p(\x\,|\,\y)\,d\x}{\displaystyle\int_{\Omega}k(\x)p(\x\,|\,\y)\,d\x
}.
 \eeq
The second part of the identity is a consequence of conservation of
matter.
%%%%%%%%%%%%%%%%%%%%%%%%%%%%%%%%%%%%%%%%%%%%%%%%%%%%%%%%%%%%%%%%
\subsection{The one-dimensional case}
%%%%%%%%%%%%%%%%%%%%%%%%%%%%%%%%%%%%%%%%%%%%%%%%%%%%%%%%%%%%%%%%
The fluxes $R_\infty$ and $R_s$ can be explicitly evaluated in
dimension 1, when the domain is a finite interval. The ratio $R_s$
was computed in \cite{KHS} in the case of an interval [0,L], when
the killing measure was uniformly distributed.

We assume now that the inward flux at $x=0$ is a constant $\Phi$ and
at $x=L$ an absorbing boundary condition is given. We consider here
the case where the killing is a Dirac $k(x)=k\delta(x-x_1)$, located
at a single point $x_1$ and $k$ is a constant. The particles are
only driven by diffusion, so the steady state equation (\ref{FPEps})
becomes
 \beqq
D\frac{\p^2p(x)}{\p x^2 }-k(x)p(x)&=&0\quad\mbox{for}\quad0<x<L \\
&&\\
\frac{\p p(L)}{\p x }&=&\Phi\\
&&\\
 p(0)&=&0
 \eeqq
and the ratio is
 \beq
R_s =\frac{D\displaystyle\frac{\p p }{\p x }(0 )}{kp(x_1)}.
 \eeq
From an explicit computation of $p(x)$, one can derive that
 \beqq
Dc'(L)&=&-\frac{D\Phi}{1+\displaystyle\frac{k}{D}(L-x_1)}\nonumber\\
&&\\
kp(x_1)&=&\frac{k\Phi(x_1-L)}{1+\displaystyle\frac{k}{D}(L-x_1)}\nonumber\\
 \eeqq and
  \beq
R_s =\frac{D}{k(L-x_1)}.\label{rss}
 \eeq
The result can be generalized to the case of a two hot spots in a
straightforward manner.
% but we present a similar computation in a
%different context in Section \ref{examples}.
When killing occurs uniformly, the ratio $R_s$ decays as  function
of $L$ like the function $1/\cosh(cL)$, ($c=const.$) \cite{KHS}.
This decay, compared to the decay of equation (\ref{rss}),  shows
that any redistribution of the killing affects this ratio, which is
discussed in the conclusion section.

In the same spirit, we give an explicit expression for $R_{\infty}$
in the case of a finite interval $[0,L]$, where particles are free
to leave the domain at the points 0 and $L$. Initially, we assume
that the particles are located at a point $x_1$. Here the killing
occurs at the point $y<x_1$. In that case the Green function
$G(x\,|\,y)$, defined by (\ref{FPEpp}), is the solution of
 \beqq
-\delta(x_1)&=&D\frac{\p^2}{\p x^2 } G(x\,|\,y)-k(x)G(x\,|\,y)
\quad\mbox{for}\quad x,y\in [0,L]\\
&&\\
G(0\,|\,y)&=&0\\
&&\\
G(L\,|\,y)&=&0
 \eeqq
and by solving this equation, the ratio $R_{\infty}$ is given by
 \beqq
R_{\infty} =\displaystyle
\frac{(L-x_1)k}{D\left(1+\displaystyle\frac{L-y}{y}-k\displaystyle\frac{y-x_1}{D}\right)}.
\eeqq
%%%%%%%%%%%%%%%%%%%%%%%%%%%%%%%%%%%%%%%%%%%%%%%%%%%%%%%%%%%%%%%%
\section{Conclusions, applications, and perspective}
%%%%%%%%%%%%%%%%%%%%%%%%%%%%%%%%%%%%%%%%%%%%%%%%%%%%%%%%%%%%%%%%
We have provided in this paper a general mathematical framework to
compute the distribution of ``killed'' and ``absorbed'' particles,
after they flow into a bounded domain. The ratios  $R_\infty$ or
$R_s$ of ``killed'' to ``absorbed'' particles are in general
difficult quantities to estimate analytically. However, in one
dimension the exact dependency of the ratio on the geometry can be
computed; we analyzed here two extreme distributions: a uniform
distribution and  a Dirac killing measure. Formulas (\ref{rss}) and
$1/\cosh(cL)$, ($c=const.$) of \cite{KHS} prove that the ratios
depend on the killing distribution. For a general three-dimensional
domain,
 $R_\infty$ can only be estimated in asymptotic cases, where  the absorbing boundary
occupies a small portion of the boundary or when the support  of the
killing measure is small (see \cite{HS}).

In the general context of microstructures in biological systems, the
ratio $R_\infty$ provides information about the total distribution
of particles. When the killing measure is redistributed and a
critical value of the ratio $R_\infty$ is attained, new biophysical
processes can be initiated that affect irreversibly the
physiological properties of the microstructure. Indeed, if enough
particles enter the structure and stay sufficiently long, they bind
to a large number of molecules. When a critical number of bonds are
made, a cascade of chemical reactions is initiated. Thus a threshold
can be reached by simply redistributing the killing measure. The
implementation of these changes at a molecular level is yet to be
identified. The mean conditional time of being absorbed before
killing, $E(T,\tau<T)$, reveals not only the time spent  inside the
structure, but also how long it takes on the average for particles
to arrive to a specific compartment.

The spine-dendrite communication can be described in terms of
quantities such as $R_\infty$ and $E(T,\tau<T)$. First, the
regulation of calcium ions that reach the dendrite can be achieved
by various mechanisms. One possibility to decrease $R_\infty$  is to
increase the length of the neck, which really occurs in vitro
experiments \cite{KS}. In that case, if the distribution of the
killing measures is scaled with the dilation of the neck, the ratio
$R_\infty$ changes, with no need to change the total killing measure
(e.g., the number of pumps). A second possibility is to redistribute
the killing measure in a way that affects the ratio $R_\infty$, as
shown in our computations (e.g., from uniform to accumulation at a
hot spot). We can predict from expression (\ref{rss}), that moving
all the calcium pumps at the bottom of the spine neck reduces the
number of ions arriving at the dendrite. Finally, the number of
pumps can also be changed. All possibilities are expected to be
observed and any particular choice should be understood in the
context of its function. We expect that the distribution of pumps
across the spine neck to be highly dynamic and driven by the mean
electrical activity of the dendrite. In particular, we may wonder
how such distribution changes in the wake of applying protocols such
as LTP (Long Term Potentiation), which lead to long term changes at
the synapse level \cite{Nicoll}. No results seem to be known about
the effect of LTP on  the pump redistribution in spines. In reality,
as studied in \cite{HSK}, the movement of ions inside the spine neck
is not purely Brownian, but has a drift component, which affects the
dynamics and changes the ratio.

The mean time $E(\tau\,|\,\tau>T)$ to arrive at the dendrite was
used in \cite{KHS} to confirm that calcium ions arriving at the
dendrite originate at the spine head (not in external sources). This
result is derived by comparing the experimental time scale with
$E(\tau\,|\,\tau>T)$. The mean time $E(\tau\,|\,\tau>T)$ is thus a
fundamental parameter in the context of spine-dendrite
communication, because it measures  the mean time calcium ions enter
the dendrite, and is  related to the induction time of cascades of
reactions, involved in modifying the synaptic weight. Changing the
$E(\tau\,|\,\tau>T)$ is a part of the spine regulation process. This
can be achieved by various ways: changing the spine neck length,
changing the number of pumps and their distribution. Various
biological investigations (see for example \cite{Sabatini}) are
dedicated to the elucidation of how such regulation is achieved at a
biochemical level.

Finally, the present computations assume that the neck width is
small. If this is not the case, the one-dimensional approximation of
the cylinder is no longer valid and pumps become insignificant.
%From a theoretical point of view, many questions remains unsolved
%concerning the survival probability in a one dimension: we wonder
%if one can  find a killing measure that leads to a new decay law?
%What is the reason that for the same decay rate for uniform killing
%on the positive axis and with a Dirac absorption?
%Another possible direction is to generalize this approach to
%include nonlinear terms. In particular, what is the asymptotic
%behavior of the survival probability when the flux condition is
%$J_{+}-J_{-} =-Ku^{q}$, where $q>1$. That is, the killing rate
%becomes a nonlinear function of the density?
%Finally, how to generalize the previous results to higher
%dimensions? In three dimensions the Brownian motion is not
%recurrent, and a Dirac distribution at a point is not enough to
%absorb as much as it does in one dimension. We are lead to the
%question of how to classify the Borel sets of a complete manifold,
%such that the survival probability has the same asymptotic decay
%law, when the killing measure is supported in these sets?

%%%%%%%%%%%%%%%%%%%%%%%%%%%%%%%%%%%%%%%%%%%%%%%%%

\end{document}